\documentstyle[11pt]{article}

\newcommand{\half}{{\scriptstyle{\frac{1}{2}}}}

\newcommand{\del}{{\mbox{\boldmath $\nabla$}}}
\newcommand{\BE}{\begin{equation}}
\newcommand{\EE}{\end{equation}}
\newcommand{\BA}{\begin{eqnarray}}
\newcommand{\EA}{\end{eqnarray}}
\newcommand{\vol}{{\sf V}}
\newcommand{\num}{{\sf N}}
\begin{document}

\begin{titlepage}

\vspace*{1mm}
\begin{center}

            {\LARGE{\bf Gravitational forces from \\ 
                 Bose-Einstein condensation  }}

\vspace*{14mm}
{\Large  M. Consoli }
\vspace*{4mm}\\
{\large
Istituto Nazionale di Fisica Nucleare, Sezione di Catania \\
Corso Italia 57, 95129 Catania, Italy}
\end{center}
\begin{center}
{\bf Abstract}
\end{center}

The basic idea that gravity can be a long-wavelength effect
{\it induced} by the peculiar ground state
of an underlying quantum field theory leads to 
consider the implications of spontaneous symmetry breaking through an 
elementary scalar field. We point out that
Bose-Einstein condensation implies the existence of long-range order and of
a gap-less mode of the (singlet)
Higgs-field. This gives rise to a $1/r$ potential and couples with infinitesimal
strength to the inertial mass of known particles. If this is
interpreted as the origin of Newtonian gravity one finds a natural solution 
of the hierarchy problem. 
As in any theory incorporating the Equivalence Principle, 
the classical tests in weak gravitational fields are fulfilled
as in general relativity. On the other hand, our picture suggests that 
Einstein general relativity may represent the weak field approximation of
a theory generated from flat space with a sequence of conformal transformations.
This explains naturally the absence of a {\it large} cosmological constant from
symmetry breaking. Finally, 
one also predicts new phenomena that have no counterpart in 
Einstein theory such as typical `fifth force' deviations 
below the centimeter scale or further modifications at distances
$10^{17}$ cm in connection with the Pioneer anomaly and the mass
discrepancy in galactic systems.

\end{titlepage}
 
\setcounter{page}{1}

\vfill
\eject
\setcounter{equation}{0}
\section{Introduction}

The basic idea that gravity is a semi-classical, 
long-wavelength effect {\it induced} 
by an underlying quantum field theory is now more than twenty years old 
\cite{fuji,zee,adler}. This approach is very appealing since, in fact, one can
get a picture of the world with only {\it three}
 elementary interactions and where
the origin of the fourth, gravitation, has to be searched in the structure of the 
vacuum. This idea has been recently re-proposed in ref.\cite{siringo} 
where the possible 
origin of gravity has been traced back to the existence of a gap-less
mode of the (singlet) Higgs field. In this paper, we shall address the basic 
problem again from scratch to make 
clear the simple physical motivations of the proposal and discuss further 
possible implications.

In the framework of induced-gravity theories, it is natural to investigate the 
possible role of spontaneous symmetry breaking. Indeed, in the Standard 
Electroweak Theory, this sets up the ground state and
is the origin of the known particle masses. 

What kind of minimal
requirements have to be met in order to obtain a consistent phenomenological
picture ? One possibility is that, hidden in some 
corner of the theory, there is
a gap-less mode of the (singlet) Higgs field that
gives rise to the attractive $1/r$ Newton
potential. It turns out that this effect, 
missed so far, can be deduced from very general 
properties such as the long-range order associated with Bose-Einstein 
condensation and the non-relativistic energy spectrum 
of low-density Bose systems with short-range two-body interactions. 

In this scenario, 
this long-range mode is then coupled in an universal way to the 
{\it inertial} mass of the known elementary fermions thus automatically
implying that `inertial mass = gravitational mass'. 
Moreover, at long distances the Higgs coupling to the fermion masses
is renormalized into the coupling to the trace of the energy-momentum tensor,
that represents the Lorentz-invariant definition of inertia. 
In this way, one can understand the origin of the
Newton constant $G$ out of a theory that, apparently, 
has only one dimensionful quantity, namely the Fermi constant $G_F$, thus 
obtaining a natural solution of the so called `hierarchy' problem. 

Finally, for weak gravitational fields, the classical tests of general 
relativity would actually be fulfilled \cite{schiff} in any theory that 
incorporates the Equivalence Principle 
and do not necessarily require an underlying {\it fundamental}
tensor theory. While
this last remark is essential for the consistency of any theoretical framework
with well known experimental results, one also predicts
new phenomena that have no counterpart in Einstein theory. For instance, 
the Newton $1/r$ potential turns out to be
modified below the centimeter scale, with possibly important consequences
for the gravitational clustering of matter. 

If, on one hand, the very accurate equality between the inertial and
gravitational mass of known particles makes, by itself, extremely natural
the idea of a `Higgs-gravity connection' \cite{bij}
to a closer inspection a tight link between 
the physical origin of gravity and the physical
origin of inertia is also suggested by classical
general relativity. Only in this case, in fact, one can understand
Einstein's formulation of the `Mach Principle', namely the consistent
vanishing of inertia if gravity would be switched off \cite{pauli}.
We shall return to this important point in the following.

The plane of the paper is as follows. In Sect.2 we shall first review the 
general features of the excitation spectrum in 4-dimensional
quantum field theories that possess a non-trivial vacuum. In Sect.3 we shall 
present the basic ingredients of spontaneous symmetry breaking in 
$\lambda\Phi^4$ theories. In Sect.4 we shall discuss the origin of the gap-less 
mode of the Higgs field. This will be shown in Sect.5 to give rise to the 
Newton potential. After, in Sect.6 we shall discuss the
connections with Einstein general relativity
in weak gravitational fields. 
Finally, we shall present in Sect.7 the summary
of our results 
together with some speculations on further possible implications at the
astronomical level.

\setcounter{equation}{0}
\section{The vacuum and its excitation spectrum}

To introduce gravity, some type
of deviation from {\it exact} Lorentz-covariance has to be introduced in order
not to run into self-contradictory statements \cite{misner}. As anticipated, 
our main point is that this type of deviation is found in the 
long-wavelength excitation spectrum of the Higgs field in the spontaneously
broken phase. Before addressing any specific detail, 
let us consider the more general aspects related to our proposal. For 
instance, the nature of the ground state may lead to violations of causality.

Quite independently of any application to gravity, 
the possible departure from an exactly Lorentz-invariant vacuum was considered
by Segal \cite{segal} as a general feature of 4-dimensional non-linear
quantum field theories, such as $(\lambda\Phi^4)_4$.
The connection with causality can be 
easily understood since the usual normal-ordering procedure guarantees the 
local commutativity of Wick-ordered products of the field operator in the free
theory. However, 
no such a procedure is known {\it a priori} for the interacting case.
Thus the argument is circular since the proper
normal-ordering procedure is only known {\it after}
determining the vacuum and its excitation spectrum.
For actual calculations, one uses the normal-ordering definition of 
free-field theory and introduces an ultraviolet cutoff $\Lambda$ or a lattice
spacing $r_o \sim 1/\Lambda$ for the 
remaining divergences \cite{zimmerman}. In this approach, where
the continuum theory is defined for
$\Lambda \to \infty$, consistency requires that the physical
spectrum should approach a Lorentz-covariant form 
$\sqrt{k^2 + M^2}$ and causality be recovered. On the other hand, 
for {\it finite} $\Lambda$, however large, 
one is faced with deviations from a Lorentz-covariant energy spectrum 
and violations of causality. 

A possible obiection is that this conclusion reflects
the use of a non-Lorentz-invariant ultraviolet
regulator. For instance, by using dimensional regularization, where the 
continuum limit is $d \to 4$, such problems should not arise. This is not so
obvious since, up to now, 
dimensional regularization is known as an essentially perturbative procedure
that, indeed, is extremely useful in those situation where a perturbative 
picture is known to work. After all, this
is the reason why one pays so much attention to the results of lattice 
simulations performed with toy-actions that only 
asymptotically possess the same symmetry properties of their 
continuum versions. At the same time, beyond perturbation theory and
just in the case of $\lambda \Phi^4$
theories, it is known that the limit $d \to 4$ is ambiguous 
\cite{autonomous} depending whether $d=4-\epsilon$ or $d=4 + \epsilon$ (for 
$\epsilon > 0$). Outside of the perturbative domain, 
similar type of problems can arise in any
theory depending on the given trajectory chosen in the complex plane to 
approach the value $d=4$. 

The problem of the excitation spectrum
becomes unavoidable, however, if one starts to model
the world as a cutoff-regulated quantum field theory since, 
in this case, the cutoff will never be removed.
However, our point of view, namely that {\it all}
departures from exact Lorentz-covariance are due to
gravitational interactions, 
offers a physical interpretation of the deviations. At the same time, 
gravity is an extremely weak interaction so that
all violations of  causality in gravitational fields
should be very difficult to observe in ordinary conditions. 
One can also reverse the argument:
if gravity is generated by the vacuum structure of a quantum
field theory and causality turns out to be effectively preserved, this means
that gravitational effects {\it cannot} become too strong. At the same time,
if we are dealing with the same physical theory, 
we would expect the problem of causality to occur in
general relativity as well. This is precisely what happens since, 
regardless of the quantum phenomena that give rise to the 
ground state, it is known 
that constant energy-density solutions
of Einstein equations contain indeed closed time-like curves \cite{godel}. 

Finally, any description of gravity should provide an explanation
of the Equivalence Principle. If Einstein theory is considered the fundamental
description of gravity, this has the
role of a  (`philosophical' \cite{baryshev1}) principle. On the other hand,
if Newtonian gravity is generated by the vacuum structure of a quantum 
field theory, it is a dynamical consequence and represents
the weak-field remnant of an otherwise exactly Lorentz-covariant 
theory. We shall return to this important point in Sect. 6.

\setcounter{equation}{0}
\section{Spontaneous symmetry breaking in $\lambda\Phi^4$ theories}

Before addressing the problem of the energy spectrum of spontaneously broken 
$\lambda\Phi^4$ theories , we have first to
consider those general properties of the
phase transition that are essential for any further analysis.

The `condensation' of a scalar field, i.e. the transition 
from a symmetric phase where $\langle \Phi \rangle=0 $
to the physical vacuum where
$\langle \Phi \rangle \neq 0  $, has been traditionally described as an
essentially classical phenomenon (with perturbative quantum 
corrections). In this picture, one uses a classical potential 
\BE
\label{clpot}
    V_{\rm cl}(\phi)= {{1}\over{2}} m^2 \phi^2 + 
{{\lambda}\over{4!}}\phi^4  
\EE
where the phase transition, as one varies the $m^2$ parameter, is 
second order and occurs at $m^2=0$. 

As discussed in ref.\cite{mech},  the 
question of vacuum stability is more subtle in the quantum theory. Here, 
the starting point is the Hamiltonian operator
\BE 
H = \, : \! \int \! d^3 x \left[ \frac{1}{2} \left( \Pi^2 + 
(\del \Phi)^2 + m^2 \Phi^2 \right) + 
\frac{\lambda}{4!} \Phi^4 \right] \! : \; 
\EE
after quantizing the scalar field $\Phi$ and the canonical momentum $\Pi$
 in terms of annihilation and creation operators
$a_{\bf k}$, $a^{\dagger}_{\bf k}$
 of a reference vacuum state $|o\rangle$
($a_{\bf k}|o\rangle=\langle o|a^{\dagger}_{\bf k}$=0).
These satisfy the commutation relations 
\BE
[ a_{\bf k}, a^{\dagger}_{{\bf k}'} ] = \delta_{{\bf k},{\bf k}'}.
\EE
and, due to normal ordering,
the quadratic part of the Hamiltonian has the usual form 
($E_k=\sqrt{ k^2 + m^2}$ )
\BE
H_2 = \sum_{\bf k} E_k a^{\dagger}_{\bf k} a_{\bf k}.
\EE
for the elementary quanta of the symmetric phase (`phions'). 

Now the trivial vacuum $|o\rangle$
 where $\langle\Phi\rangle=0$ is clearly locally 
stable if phions have a physical mass $m^2 >0$.
  However, is an $m^2>0$ 
symmetric vacuum necessarily {\it globally} stable ?  Could the phase 
transition actually be first order, occurring at some small but positive 
value of the physical mass squared
$m^2 > 0$? The question is not entirely trivial just because \cite{mech} 
the static limit of the
2-body phion-phion interaction is not always repulsive. Besides the
tree-level repulsive potential
there is an induced 
attraction from higher-order graphs. In this case, 
for sufficiently small values of $m$, the trivial `empty' state  $|o\rangle$ 
may not be the physical vacuum.

  The answer to the question depends on the form of the 
{\it effective potential} $V_{\rm eff}(\phi)$ 
and it is not surprising that different approximations
may lead to contradictory results on this crucial issue. The situation is 
similar to the Bose-Einstein condensation in condensed matter
that is a first-order 
phase transition in an ideal gas. However, 
in interacting systems the issue is more delicate and often difficult to be
settled experimentally. Theoretically is predicted to be 
a second-order transition in some approximations but it may appear as
a weak first-order transition in other approximations \cite{trento}. 

We shall refer to \cite{mech,zeit} for details on the 
structure and the meaning of  various types of
approximations to the effective
potential and just report a
few basic results:

\par~~~~i) the phase transition 
 is indeed first order as in the case of the simple one-loop potential. This
is easy to realize 
if one performs a variational procedure, within a simple
class of trial states that includes $|o\rangle$. In this case, one
finds \cite{cian} that the $m=0$ theory lies in the broken phase. Therefore
the phase transition occurs earlier, for some value of the phion mass 
$m\equiv m_c$ that is still positive.
 This conclusion is confirmed by the results 
of ref.\cite{rit2} that provides the most accurate non-perturbative
calculation of the 
effective potential of $\lambda\Phi^4$ theories performed so far. 

Understanding the magnitude of $m_c$ requires additional comments. 
As recalled in Sect.2, 
the normal ordering prescription in Eq.(3.2) eliminates all ultraviolet 
divergences of the free-field case at $\lambda=0$. However, 
for $\lambda >0$ there are additional divergences. For this reason, one
introduces an ultraviolet cutoff $\Lambda$ and defines the continuum 
theory as a suitable limit $\Lambda \to \infty$.  In this case, however, 
one is faced with a dilemma since a meaningful description of
SSB in quantum field theory {\it must} provide $m_c=0$.
Otherwise, from the existence of a non-vanishing mass gap controlling
the exponential decay of the two-point function of the symmetric phase, and
the basic axioms of quantum field theory \cite{glimm}
 one would deduce the uniqueness of the 
vacuum (and, thus, no SSB). The resolution of this apparent conflict 
\cite{zeit,ritschel,mech} is that the continuum limit 
of the cutoff-regulated theory gives a vanishing ratio 
($\epsilon\equiv{{1}\over{\ln {{\Lambda}\over{M_h}} }}$)
\BE
{{m^2_c}\over{M^2_h}} \sim \epsilon 
\EE
so that when the scale of the spontaneously broken phase, namely
the Higgs boson mass $M_h$, is taken as the unit scale of mass,
the possible values of the phion mass $0 \leq m \leq m_c$ are
 naturally infinitesimal. 
In this sense, SSB is an
 {\it infinitesimally weak} first-order phase transition where
the magnitude of the ratio
${{m}\over{M_h}}$ represents a measure of the degree of
non-locality of the cutoff-regulated theory.

\par~~~~ii) there is a deep difference between a `free-field' theory and a
`trivial' theory \cite{book} where the interaction effects die out in the
continuum limit. The former has a quadratic effective 
potential and a unique ground state. The latter, even for a vanishingly
small strength $\lambda={\cal O}(\epsilon)$ of the elementary two-body 
processes can generate a {\it finite} gain in the energy density, and thus
SSB, due to the macroscopic occupation of the same
quantum state, namely to the phenomenon of Bose condensation. This leads to
a large re-scaling of $\langle \Phi \rangle$. Indeed, one can 
introduce, in general, 
two distinct normalizations for the vacuum field $\phi$, say a `bare' field
$\phi=\phi_B$ and a `renormalized' field
$\phi=\phi_R$. They are defined through the quadratic 
shapes of the effective potential in the symmetric and broken phase 
respectively
\BE
\left. \frac{ d^2 V_{\rm eff}}{d \phi_B^2} \right|_{\phi_B=0} \equiv  m^2, 
\quad \quad \quad 
\left. \frac{ d^2 V_{\rm eff}}{d \phi_R^2} \right|_{\phi_R=v_R} \equiv M_h^2.
\EE
Due to `triviality', the theory is ``nearly'' a massless, free theory so that
 $V_{\rm eff}$ 
is an extremely flat function of $\phi_B$. Therefore, due to (3.5), 
the re-scaling $Z_\phi$ relating $\phi_B$ and $\phi_R$ becomes very large.
 By defining $\phi^2_B = Z_\phi \phi^2_R$, one finds 
$Z_{\phi} ={\cal O}( {{1}\over{\epsilon}} )$ or
\BE
           v_R \sim v_B \sqrt{\epsilon}
\EE
Just for this reason,
 the rescaling of the `condensate' $Z=Z_\phi$ is different 
from the more conventional quantity $Z=Z_{\rm prop}$ defined from the 
residue of the shifted field propagator at $p^2=M^2_h$. According to 
K\"allen-Lehmann decomposition and `triviality' this has
a continuum limit $Z_{\rm prop}=1 +{\cal O}(\epsilon)$.

\par~~~~iii) the existence of
two different continuum limits $Z_\phi \to \infty$ and
$Z_{\rm prop}\to 1$  reflects a fundamental discontinuity in the
2-point function at $p=0$ ($p$= Euclidean 4-vector). This effect is not
totally unexpected and its origin 
should be searched in the infrared divergences of 
perturbation theory for 1PI vertices at zero external momenta \cite{syma}.
Of course, after the Coleman-Weinberg \cite{cw} analysis, we know how to obtain 
infrared-finite expressions for 1PI vertices at zero external momenta. This
involves summing up an infinite series of graphs of different perturbative
order with different numbers of external legs, just as in the analysis of 
the effective potential that was taken as the starting point for our analysis.
In this case, the second derivative of the effective potential gives 
$\Gamma^{(2)}(p=0)$,
the inverse susceptibility
\BE
\chi^{-1}=
\left. \frac{ d^2 V_{\rm eff}}{d \phi_B^2} \right|_{\phi_B=v_B} 
={{M_h^2}\over{Z_\phi}}
\EE
Therefore, if
$Z_{\phi} ={\cal O}( {{1}\over{\epsilon}} )$, one finds
\BE
         {{\Gamma^{(2)}(0)}\over{M^2_h}} \sim \epsilon 
\EE
rather than $\Gamma^{(2)}(0)= M^2_h$ as expected for a free-field theory where
\BE
\Gamma^{(2)}(p) = (p^2 + M^2_h)
\EE
Notice that the discrepancy found in the discrete-symmetry case implies the
same effect for the
zero-momentum susceptibility of the {\it radial} field 
in an O(N) continuous-symmetry theory. This conclusion, besides the
general arguments of \cite{syma}, is supported by the explicit calculations
of Anishetty et al \cite{ani}.

Notice that SSB requires the subtraction of disconnected pieces so that
continuity at $p=0$ does not hold, in general \cite{plb}. At the same time, 
a mismatch at $p=0$ does not violate `triviality' since no scattering 
experiment can be performed with exactly zero-momentum particles.
On the other hand, for large but
finite values of the ultraviolet cutoff $\Lambda$, when `triviality' is not
complete, the discrepancy between 
$\Gamma^{(2)}(0)$ and $M^2_h$ will likely
`spill over' into the low-momentum region 
$p^2 \sim \epsilon M^2_h$. In this region, we expect
sizeable differences from the free-field form Eq.(3.10). 

If really $Z_\phi \neq Z_{\rm prop}$ this result has to show up in
sufficiently precise numerical simulations of the broken phase.
To this end, 
the structure of the two-point function has been probed in refs.
\cite{cea} by using the largest lattices considered so far.
 One finds substantial deviations from Eq.(3.10) in the low-$p$ region and
 only for large enough $p$, $\Gamma^{(2)}(p)$ approaches
the free field form (3.10). Also, 
the lattice data of refs.\cite{cea}
support the prediction that the 
discrepancy between 
$ \Gamma^{(2)}(0)$ 
and the asymptotic value 
$M^2_h$ becomes larger when approaching
the continuum limit. 

We stress that 
no such a discrepancy is present in the symmetric phase where 
$\langle \Phi \rangle=0$. Here the free-field behaviour 
$ \Gamma^{(2)}(p) = (p^2 + m^2)$ is valid to high accuracy
down to $p=0$ \cite{cea}. Notice the 
different effect of the cutoff in the broken and symmetric phases. In both 
cases, the limit $\Lambda \to \infty$ yields a free spectrum of the type
(3.10). In the broken phase, however, this is obtained 
by making sharper and sharper 
a discrepancy at low $k$ so that
a discontinuity at $k=0$ will remain.

In conclusion: theoretical arguments and numerical evidences suggest that
in the limit $k \to 0$
the excitation spectrum of the broken phase 
can show substantial deviations from the free-field form
$\tilde{E}=\sqrt{k^2+M^2_h}$. Due to the `triviality' of the theory, 
the deviations from the free-field behaviour should, however,
be confined to a range of $ k$ that becomes infinitesimal
in units of $M_h$ in the continuum limit $\Lambda \to \infty$. 

\setcounter{equation}{0}
\section{A gap-less mode of the Higgs field: the vacuum is not `empty'}

In this section we shall present some very general  arguments to illustrate
the nature of the excitation spectrum of the broken phase
in the limit $k \to 0$. The discussion is valid in the framework of 
a weakly first-order phase transition where one can meaningfully
describe the broken phase as a Bose condensate of the elementary quanta of the
symmetric phase.

The starting point for our analysis is 
a positive-definite (but otherwise arbitrary) relation
between the number density $n$ of condensed phions at $k=0$
and the scalar field expectation value, namely
\BE
n = n(\phi_B)
\EE
Using Eq.(4.1) one can easily transform
the energy density ${\cal E}={\cal E}(n)$ into the effective potential 
$V_{\rm eff}=V_{\rm eff} (\phi_B)$. In this way, 
the $\phi_B=0$ `mass-renormalization' condition
in (3.6) 
\BE
\left. \frac{ d^2 V_{\rm eff}}{d \phi_B^2} \right|_{\phi_B=0} \equiv  m^2
\EE
becomes
\BE
\left. \frac{ \partial {\cal E}}{\partial n} \right|_{n=0} = m.
\EE
Its physical meaning is transparent. If we
consider the symmetric vacuum state 
(``empty box'') and  add a very small density $n$ of phions (each with 
vanishingly small
3-momentum $k \to 0$)
the energy density is 
$ {\cal E}(n) - {\cal E}(0) \sim nm$ 
in the limit $n \to 0$.

Let us now analyze spontaneous symmetry breaking. This can be viewed as a 
phion-condensation process occuring at those values $\phi_B=\pm v_B$ 
where 
\BE
\left. \frac{ d V_{\rm eff}}{d \phi_B} \right|_{\phi_B=v_B}=0
\EE
By using Eq.(4.1) and defining the ground-state particle density
\BE
n_v = n( v_B)
\EE
we also obtain
\BE
\left. \frac{ \partial {\cal E}}{\partial n} \right|_{n=n_v} = 0
\EE
Eq.(4.6), differently from Eq.(4.3), means that small changes of the phion density around 
its stationarity value 
do not produce any change in the energy density of the system. Namely, 
$ {\cal E}(n)-{\cal E}(n_v) \sim (n-n_v)^2$ and, as a consequence of 
condensation, one can add or remove an arbitrary number of
phions at $k = 0$ without any energy cost, just as in the non-relativistic
limit of the theory. 
\vfill
\eject
Therefore, for $ k \to 0$, 
the excitation spectrum of the theory exhibits the following 
features:
\vskip 10 pt
~~~~a) {\underline {in the symmetric phase}}~~~~~~~~~~~~ $E(k) \sim m + 
{{k^2}\over{2m}} \to m $ 
\par (which is the standard spectrum for massive particles)
\vskip 10 pt
~~~~b) {\underline{in the broken phase}}~~~~~~~~~~~~~~~~~~$\tilde{E}(k)\to 0 $  
\par ( which is the condition of a gap-less spectrum) 
\vskip 10 pt
In this sense, in the broken phase, the $k \to 0$
Fourier component of the scalar field 
behaves as a massless field.  We now understand why, in the broken phase,
the excitation spectrum $\tilde{E}$ 
cannot be $\sqrt{k^2 + M^2_h}$ when $k \to 0$: this 
form does not reproduce $\tilde{E}=0$ at $k=0$. 

One may object
that the excitation spectrum is {\it discontinuous} so that one has exactly
$\tilde{E}=\sqrt{k^2 + M^2_h}$ for {\it all} $k\neq 0$ except at $k=0$. 
First of all,
this is not what is generally 
believed since the rest mass $M_h$ is generally identified with the energy-gap
$\tilde{E}(k=0)$
of the broken phase (at least in the discrete-symmetry case where 
there are no Goldstone bosons).
Moreover, as anticipated in Sect.3, this type of
behaviour is, indeed, expected for the {\it continuum} theory. 
In a cutoff theory, 
however large $\Lambda$ may be, all singularities are smoothed and
one has a continuous spectrum for all values of $k$. 
This point of view is also consistent with the sequence of
lattice calculations of ref.\cite{cea}.

The existence of the gap-less mode is directly related to
Bose-Einstein condensation 
\cite{trento,anderson}. Indeed, the
behaviour of the spectrum for $k \to 0$ in the broken phase
corresponds to the range of momenta $k \ll m$ of the {\it non-relativistic} 
theory. In this regime a scalar condensate, whatever its origin may be,
is a highly correlated structure with {\it long-range} order. 
This is clear from the following general argument \cite{trento} due to Anderson.
Suppose that in a large box of volume
$\vol \to \infty$ we have a condensate of
$\num\to \infty $ particles in the $k=0$ mode. Let us divide the box into a 
large number of $K$ identical and
macroscopic subsystems so that each subsystem still 
contains a very large number 
$(\num/K)$ of particles. Let us also denote $A_i$ the annihilation 
operator for the $k=0$ particles contained in the $i$th subsystem. In this
case, we have
$\langle a^{\dagger}_o a_o\rangle=\num$, 
$\langle A^{\dagger}_i A_i\rangle=\num/K$, so that from
\BE
          a_o= {{1}\over{ \sqrt{K} }} \sum_i A_i
\EE
we obtain
\BE
           \num={{\num}\over{K}} + 
{{1}\over{K}} \sum_{i \neq j} \langle A^{\dagger}_iA_j \rangle 
\EE
For large $K$, the second term must dominate so that, the phases of the $A_i$ in
 different  subsystems must be correlated. In the limit where $K \to \infty$ 
(but still $\num/K$ is a large number)
this is equivalent to
introduce a {\it complex} condensate wave-function at each point in space
$\Psi \sim \sqrt{  n } e^{i \theta }$ that represents the true order parameter
to describe the response of the condensate to the very long wavelengths with
$k \ll m$. Although the energy density does not depend on the possible
constant values of the phase, the vacuum state will pick up
just one of them. In this sense, 
the gap-less mode of the Higgs field can be considered 
the Goldstone boson of a spontaneously broken {\it continuous}
symmetry, the phase rotations of the non-relativistic wave-function of the 
condensate that does not exist in the symmetric phase. 

Therefore, even in the case of spontaneous symmetry breaking with
a {\it neutral} scalar field, the ground state 
is still infinitely degenerate \cite{araki} and
adding particles with $k \to 0$ will only induce
an energy density $ \sim ({\bf {\nabla}}\theta)^2$. Truly enough, 
this effect shows up only for very small values of $k$
and is not perceivable at short distances. Just for this reason, 
the peculiar phenomenon at the basis
of the gap-less mode of the Higgs field has nothing to do with the 
Goldstone bosons that give rise to the $W$ and $Z$ masses. These arise from the
spontaneous breaking of continuous symmetries that are already seen in the 
symmetric phase (where there are no condensates whatsoever).

Notice that our  result, 
although deduced within the framework of ref.\cite{mech}, 
does not depend on the validity of the relation
\BE
n \sim \half m \phi^2_B,
\EE
used in ref.\cite{mech}.
 Indeed, Eq.(4.6) follows
from Eq.(4.4) {\it regardless} of the precise functional relation between 
the phion density $n$ and the vacuum field $\phi_B$. 
Moreover, the same conclusions hold 
in any description based on a first-order phase transition
where the broken phase can be represented
as a condensate of the elementary quanta of the symmetric phase.
For instance, the phase transition remains
first-order if spontaneous symmetry breaking is induced by
(or contains the additional contributions of) intermediate vector bosons 
\cite{cw,ibanez}. In this case, as in pure $\lambda\Phi^4$ theory, the
massless theory at $m=0$ is found in the broken phase so that the phase
transition occurs earlier at a non-zero and positive $m^2_c$.

After this general discussion, let us now attempt
a semi-quantitative description of the energy spectrum of the broken phase.
A first observation is that  
for $k\sim m$  (or larger) we expect
\BE
\tilde{E}(k) \sim \sqrt{k^2 +M^2_h}
\EE
On the other hand, the region $k \to 0$ 
of  {\it low-density} Bose systems, with short-range 2-body interactions, 
can be analyzed in a universal way \cite{universal} namely
\BE
     \tilde{E} \sim c_s k~~~~~~{\rm for}~k\to 0 
\EE
where $c_s$ is the sound velocity. 
In a simple picture, the two branches of the spectrum 
join through some form of continuous 
matching at momenta $k \sim m$ (see fig.1), 
analogously to the case of `phonons' and `rotons' in superfluid He$^4$. 
 If we recall
that $M_h >> m$, we then obtain the order of magnitude estimate
\BE
           c_s \sim {{M_h}\over{m}}
\EE
This result can be expressed in a more quantitative form if one uses
the precise relation for dilute Bose systems \cite{huang}
\BE
            c_s \equiv {{1}\over{m}} \sqrt{4\pi n a}
\EE
in terms of
the physical S-wave `phion-phion' scattering length $a$. In this case, 
by using the results
of ref.\cite{mech}, we find
$a\sim {{\lambda}\over{8\pi m}} $ and
the expression for the Higgs mass \cite{mech}
\BE
M^2_h \equiv 8\pi n a
\EE
so that we get the final result 
\BE
             c_s = {{M_h}\over{m \sqrt{2} }} \equiv \sqrt{\eta}
\EE
Notice that Eq.(4.15) makes no reference to the bare coupling
$\lambda$ entering the hamiltonian density (3.2) and would be formally 
unchanged if the scalar self-interaction were replaced by a short-range 
interaction that includes the effect of vector-boson and/or fermion 
loops \cite{cw}.

Eqs. (4.11) and (4.13)
become a better and better approximation 
in the limit of very low-densities
$na^3 \to 0$ where all condensed phions are found in the state
at $k=0$ and there is no population of the finite momentum modes
( `depletion') since
\BE
        D\equiv 1 - {{\num (k=0)}\over{\num}} ={ \cal O} (\sqrt{na^3})
\EE
The depletion is a simple phase-space effect representing the probability
that, besides the condensate, also states such as
$({\bf{k}},-{\bf{k}})$ are populated in the ground state. 
It represents a measure 
of interaction effects that cannot be re-absorbed into the linear
energy spectrum \cite{huang} and, therefore, can be viewed as {\it residual}
interaction.
 In this respect, 
spontaneous symmetry breaking in a cutoff $\lambda\Phi^4$ theory corresponds
to the case of an {\it almost} ideal, dilute Bose system. In fact, 
`triviality' requires a continuum limit with
 a vanishing strength $\lambda={\cal O}(\epsilon)$ 
for the elementary 2-body processes. 
Together with Eq.(3.5), this leads to $aM_h \sim \sqrt{\epsilon}$. Therefore, 
 by taking $M^2_h\equiv 8\pi n a$ 
as  the physical scale of the theory in the broken phase, we find a continuum 
limit where $a \to 0$, $n \to \infty$ with $na=const.$ and
\BE
                   n a^3 ={\cal O}(\epsilon)
\EE
When $\epsilon \to 0$, the phion-condensate becomes infinitely dilute so that
the average spacing between two phions in the condensate, 
$d \equiv n^{-1/3}$, becomes enormously larger than their scattering length.
In this limit, 
the energy spectrum (4.11) becomes exact ( for $k \to 0$) while, for finite 
$\Lambda$, 
there are ${\cal O} (\epsilon)$ corrections and a small, but finite, depletion
with density 
\BE
{{n_D}\over{n}}= {\cal O}( \sqrt \epsilon)
\EE 
Notice, however, that the phion density
$n \sim {{1}\over{2}}m v^2_B$, is very large, 
${\cal O}(\epsilon^{-1/2})$, in the physical units denoted by the correlation
length $\xi_h\equiv1/M_h$. Indeed, 
${{d}\over{\xi_h}} \sim \epsilon^{1/6}$.  It is because there is such a 
high density of phions that their tiny 2-body
interactions ${\cal O}(\epsilon)$ can produce a finite effect 
on the energy density. In this sense, the phion condensate is a very dilute gas
when observed
on  the very small scale of the phion-phion scattering length $a$ but may appear as a 
very dense {\it liquid} on  larger scales. 

At the same time, Bose liquids
at zero-temperature are known to possess the remarkable property of 
{\it superfluidity} so that the scalar condensate, when placed in an external 
field, flows without friction. For this reason, 
the result $\tilde{E} (k) \sim c_s k$
for $k \ll m$, deduced 
from the quantum dynamics of weakly 
coupled Bose systems with short-range two-body interactions, 
could have been obtained by requiring
a frictionless motion of macroscopic bodies in the vacuum. Namely, 
the condition 
that the scalar condensate cannot absorb arbitrarily small amounts 
of energy-momentum transfer, for $ k \ll m$, is
precisely the starting point used by Landau \cite{landau} to deduce the linear 
excitation spectrum of a superfluid at low $k$ where the motion is
frictionless provided 
the velocity of an external body is
$ |{\bf{u_e}}| \leq c_s$. In the case of the phion condensate,
this is not a restriction in view of the fantastically high value of 
the `sound-velocity' $c_s \sim {{M_h}\over{m}}~c >> c$ \cite{feynman}. At the
same time, the residual self-interaction effects embodied in the presence
of a non-zero depletion can give rise to a small friction when studying the
superfluid flow over very large distances.

In conclusion: spontaneous symmetry breaking in a cutoff $\lambda\Phi^4$ theory
gives rise to an excitation spectrum that is {\it not} exactly 
Lorentz-covariant. The usual assumption 
$\tilde{E}(k) \sim \sqrt{k^2+M^2_h}$ 
is not valid in the limit $k \to 0$ where one actually finds a `sound-wave' 
shape $\tilde{E}(k) \sim c_s k$.
This result reflects the
{\it physical} presence of the scalar condensate. However, 
as expected from our analysis in Sect.3, all deviations from a free-field
spectrum are confined to a range of momenta $k \ll m$ that becomes 
infinitesimal, in units of $M_h$, in the limit $\Lambda \to \infty$.

\setcounter{equation}{0}
\section{A long-range potential in Higgs condensates}

It is well known that
condensed-matter systems can support long-range forces
even if the elementary constituents have only short-range 2-body 
interactions. Just for this reason, it is not surprising that
the existence of a gap-less mode for $k \to 0$ in the broken phase
can give rise to a long-range potential. For instance, when
coupling fermions to a (real) scalar Higgs field with vacuum 
expectation value $v$ through the Standard Model interaction term 
\BE
         -m_i \bar{\psi}_i \psi_i (1+ {{h(x)}\over{v}})
\EE
the static limit $\omega \to 0$ of the scalar propagator
\BE
                D(k,\omega)= {{1}\over{\tilde{E}^2(k)- \omega^2 -i0^+}}
\EE
gives rise to an attractive potential
between any pair of masses $m_i$ amd $m_j$ 
\BE
          U(r)=-{{m_im_j}\over{v^2}} 
\int {{d^3k}\over{(2\pi)^3}} 
{{ \exp (i {\bf{k}}\cdot {\bf{r}})  } \over{  \tilde {E}^2(k)   }}
\EE
By assuming Eqs.(4.10) and (4.11) for $ k \to \infty$ and $k \to 0$, 
and using
the Riemann-Lebesgue theorem \cite{goldberg} on Fourier transforms, the
leading $r \to \infty$
behaviour is universal. Any form of the spectrum that for $k \sim m$
interpolates between the two asymptotic trends would produce the same
result. At large distances $r >> 1/m$ one finds
($\eta = {\cal O} ({{1}\over{\epsilon}})$)
\BE
            U(r)=- {{G_F}\over{4\pi \eta}}{{m_im_j}\over{r}}
[1 + {\cal O}(1/mr)]
\EE
where $G_F\equiv 1/(v^2)$. 
In the physical case of the Standard Model one would identify
$G_F \sim 1.1664 \cdot 10^{-5}$ GeV$^{-2}$ with the Fermi constant.

Notice that the coupling in Eq.(5.1) naturally defines the `Higgs charge' 
of a given fermion as its physical mass. However, 
for nucleons, this originates from
more elementary Higgs-quark and quark-gluon interactions. At low $k$
these effects can be resummed to all orders by replacing Eq.(5.1) with the 
effective 
coupling to the trace of the energy-momentum tensor
$T^{\mu}_{\mu}$. Namely, by denoting $\tilde{h}$ the long-wavelength of the 
Higgs field associated with the linear part of the spectrum, we can write down
the first few terms of an effective lagrangian
($\tilde{\phi}\equiv {{ \tilde{h} (x)}\over{v}}$) 
\BE
 {\cal L} (\tilde{\phi})=  
{{v^2}\over{2}} \tilde{\phi}
       [ \eta \nabla^2 - {{\partial^2 } \over{ c^2\partial t^2}}] \tilde{\phi}
      - T^{\mu}_{\mu}(1+ \tilde{\phi})  + ...
\EE
In Eq.(5.5) the dots indicate cubic and higher order terms describing 
residual self-interaction effects of the type discussed in Sect.4 and we have 
made explicit the factor $\eta$ coming from the peculiar nature of the 
energy spectrum $\tilde{E}(k)=\sqrt{\eta}k$ for $k \to 0$. 
The different normalization of the linear coupling reduces to the
usual definition in the case of free fermions and yields exactly the nucleon mass
when evaluating the matrix element between nucleon states 
\BE
        \langle N | T^{\mu}_{\mu}| N \rangle =m_N \bar{\psi}_N \psi_N
\EE
We note that the strength of the long-range potential is proportional to the
product of the masses and is
naturally infinitesimal in units of $G_F$. It would vanish 
in a true continuum theory where the gap-less mode of the Higgs field 
disappears and $\eta \to \infty$.
Therefore, it is natural to relate this extremely weak
interaction to the gravitational potential
and to the Newton  constant $G$ by identifying
\BE
              \sqrt{\eta} = \sqrt{ {{G_F}\over{G}} } \sim 10^{17}
\EE
Notice that Eq.(5.5) rensembles a Brans-Dicke theory \cite{brans}.
Here, however, the framework is very different since the $\tilde{\phi}-$field 
propagates in the presence of the phion condensate. 
Approximating the field $\tilde{\phi}$ as
a free field with $\tilde{E}^2(k)= \eta k^2$, 
we can write down its equation of motion, namely
\BE
       [ \eta \nabla^2 - {{\partial^2 } \over{ c^2\partial t^2}}] \tilde{\phi}=
         { { T^{\mu}_{\mu} }\over{ v^2}}
\EE
that, due to the fantastically high value of $\eta$, reduces for all practical 
applications to 
\BE
          \nabla^2 \tilde{\phi}= G T^{\mu}_{\mu} 
\EE
Finally, for classical motions 
in the limit of velocities $|{\bf{u}}_n| \ll c$, when the trace of the
energy-momentum tensor \cite{weinberg}
\BE
         T^{\mu}_{\mu}(x) \equiv \sum_n 
{ { E^2_n -   {\bf{p}}_n  \cdot  {\bf{p}}_n  } \over{E_n}} 
\delta^3 ( {\bf{x}} - {\bf{x}}_n(t) ) 
\EE
reduces to the mass density
\BE
         \sigma(x ) \equiv \sum_n m_n
\delta^3( {\bf{x}} - {\bf{x}}_n(t)) 
\EE
Eq.(5.9) becomes the Poisson equation for the Newton potential 
\BE
            \nabla^2 \tilde{\phi}= G  \sigma( x)
\EE
Notice that the long-range $1/r$ potential is a direct consequence 
of the existence of the scalar condensate. Therefore, speaking of
gravitational interactions makes sense only for particles that can 
induce variations of the phion density by exciting the gap-less
mode of the Higgs field.
 In this sense, phions, although possessing an inertial mass, have
no `gravitational mass'. 

In the physical case of the Standard Model, and assuming the
range of Higgs mass
$M_h \sim 10^2-10^3$ GeV, we obtain
a range of phion masses
$m \sim 10^{-4}-10^{-5}$ eV. 
The detailed knowledge of the spectrum 
$\tilde{E}(k)$ for $k\sim m $ would allow to compute the terms
${\cal O}(1/mr)$ in Eq.(5.4) and predict a characteristic pattern of
 `fifth force' deviations below the centimeter scale. 

As anticipated in Sect.4, the nature of 
the vacuum implies the scalar condensate to behave as a superfluid at 
zero temperature. Therefore, its velocity field ${\bf {u_s}}$ corresponds
to a potential flow 
\BE
                 {\bf{\nabla} }~ {\bf{x}} ~ {\bf {u_s}}= 0
\EE
This provides a simple hydrodynamical picture of Newtonian gravity. Indeed, by
identifying
\BE
 {\bf {u}}_s= {{ {\bf {\nabla}} \tilde{\phi}}\over{m}}
\EE
and introducing an average constant phion density $\langle n \rangle$, 
the Poisson equation can be re-written as 
\BE
              {\bf {\nabla}} {\bf {\cdot}} (\langle n \rangle {\bf{u}}_s) = 
\sigma (x)
\EE
provided we identify 
\BE
\langle n \rangle= {{m}\over{ G}}
\EE
Eqs.(5.14) and (5.15) establish a formal relation between the difference 
of the gravitational potential and the associated superfluid flow. In this
picture $\tilde{\phi}$ is a classical field determining
the phase of the non-relativistic condensate wave function 
$\Psi \sim \sqrt{ \langle n \rangle} e^{i \tilde{\phi} }$ 
\cite{lifshitz}.

On the other
hand, the Poisson equation (5.12) is modified if
the phion density $n \equiv f(x) \langle n \rangle$ sizeably differs from its 
constant value (5.16) related to the Newton constant. In this case we find instead
\BE
        {\bf {\nabla}} {\bf {\cdot}} (n 
{\bf{u}}_s) =  \sigma (x)
\EE
Therefore, since the scalar density $n$ can
depend on $x$ only through the local gravitational
acceleration field, we obtain the modified Poisson equation
\BE
   {\bf {\nabla}} {\bf {\cdot}} \left[ f
({{|{\bf{ \nabla}} \tilde {\phi}| }\over{ g_o }})
{\bf{ \nabla}} \tilde {\phi} \right]
= G \sigma (x)
\EE
where a constant acceleration $g_o$ has been introduced to make
$f$ dimensionless. The transition from Eq.(5.15) to
Eq.(5.17) corresponds to include the effects 
of residual self-interactions 
into the corresponding hydrodynamics 
of a low-temperature Bose liquid (see \cite{lifshitz}). In this
case, Eq.(5.15) corresponds to the linearized approximation.

Notice that Eq.(5.18) is formally identical to the non-linear
modification of inertia ( `MOND' )
introduced by Milgrom \cite{milgrom} to resolve
the substantial mass discrepancy and describe many experimental features of
galactic systems. This approach
represents an alternative to the dark-matter hypothesis and predicts
drastic departures from Netwonian dynamics in the typical astronomical 
large-scale and low-acceleration conditions. These occur
when the gravitational acceleration of bodies
becomes comparable to a cosmic acceleration field
($H$ is the Hubble constant)
\BE
        g_o \sim cH \sim 10^{-8}~{\rm cm}~{\rm sec}^{-2}
\EE
In our picture, this should also correspond to a regime where the phion density
$n$ differs substantially from its value in (5.16).

The connection with the Hubble constant can be understood by exploring the
implications of Bose-Einstein condensation in an expanding universe. In this
case, there must be a continuous creation of phions at a rate
\BE
   {{\delta \num}\over{\num}} \sim {{\delta \vol}\over{\vol}} \sim 3H \delta t
\EE
to maintain the same particle density $n_v$ that minimizes
the energy density Eqs.(4.4)-(4.6) \cite{bondi}. This gives rise to
a cosmic flow $\langle {\bf{u}}_s \rangle$ that may be used to
define a cosmic acceleration field
\BE
       g_o \equiv  m |\langle {\bf{u}}_s \rangle|
\EE
that does not depend on the local distribution of gravitational sources.
If Hubble expansion takes place only in intergalactic space, the effects
of $g_o$ are not observable in solar-system tests.
We shall return to this point in the conclusions.

\setcounter{equation}{0}
\section{Comparison with general relativity in weak
 gravitational fields}

To illustrate the connection with general relativity, we observe
preliminarly that Einstein's description of gravity is purely geometric and 
macroscopic. As such, it does not depend on any hypothesis about
the  physical origin of this interaction. For instance, classical
general relativity, by itself, is unable to
predict \cite{pauli} even the {\it sign} of the gravitational force 
(attraction rather than gravitational repulsion). Rather, Einstein had to 
start from the peculiar properties of
Newtonian gravity to get the basic idea of transforming the classical effects of
this type of interaction into a metric structure. For this reason, 
classical general relativity cannot be considered a
truly {\it dynamical} explanation of the origin of the gravitational forces. 

It is obvious that
in a description where gravity is a long-wavelength excitation of the scalar 
condensate there are differences with respect to the standard ideas.
For instance, the gravitational force is naturally instantaneous.
The velocity of light $c$ is quite unrelated to
the long-wavelength excitations of the scalar condensate that for $k \to 0$
propagate with the fantastically high speed 
$c_s= \sqrt{\eta} c\sim 10^{17} c$.
Only for $k \sim m$, i.e. at the joining of 
the two branches of the excitation spectrum, one recovers the expected result 
$d\tilde{E}/dk < c$.  To a closer inspection, 
this apparently bizzarre result appears
less paradoxical than the generally accepted point of view that considers
the inertial forces in an accelerated laboratory as the consequence of a 
gravitational wave generated by distant accelerated matter. Indeed, if the
gravitational interaction would really 
propagate with the light velocity, distant matter
must be accelerated {\it before} the inertial 
reaction is actually needed \cite{dicke}. 
Similar conclusions are also suggested by the analysis of tidal forces 
\cite{narlikar}. 

The instantaneous nature of the long-range
gravitational interaction is a direct consequence
of its non-local origin from the scalar condensate. 
As anticipated in the Introduction, and on the basis of the hydrodynamical 
picture of gravity of Sect.5, this leads to a
`Mach's Principle' view of inertia. Indeed, one can imagine that
removing at spatial infinity 
{\it all} gravitational sources produces an
infinite flow of the scalar condensate and,as a
net result, the vacuum becomes `empty' around
a given body and its inertia vanishes \cite{remark}.
Therefore, one cannot speak of absolute accelerations 
with respect to empty space since the 
inertial mass of a test particle depends on the existence
of the scalar condensate whose density is determined by the distribution of 
gravitating matter. In this sense, the `Mach
Principle' represents a concise formulation of the inextricable 
connection between
inertia and gravity due to their common origin from the same physical
phenomenon: the condensation of the scalar field . Notice that
Mach's ideas had 
a strong influence on the origin of general relativity \cite{einstein}. 
However, to our knowledge, the physical mechanisms for which matter `there' can
determine inertia `here' had never been addressed.

Finally, one should consider the different 
impact on general relativity of possible modifications of the Newton potential. 
A long-distance replacement
${{1}\over{r}} \to {{\exp (-\mu r)}\over{r}}$ was indeed considered by Einstein
\cite{cosmo} in connection with the cosmological problem and the
introduction of a cosmological term in the field equations. On the other hand, 
a modification of the $1/r$ potential below the centimeter scale would imply 
that classical general relativity is a truly {\it effective} 
theory. As such, it would hardly make sense to consider it a `bare' theory, 
i.e. the starting point for a quantization 
procedure. While this point of view
is consistent with the induced-gravity approach, where Einstein theory
represents, indeed, the weak-field approximation in an all-order expansion in
the Riemann tensor \cite{fuji,zee,adler}, one should realize that,
if gravitational clustering of matter is modified at short distances, the
Schwarzschild singularity may be just an artifact of the approximation.

After this preliminary discussion, let us try to understand 
whether our description of gravity is consistent with the experimental 
results. We observe that our picture yields Newtonian gravity and, as such, 
predicts that bodies with different inertial masses undergo the same 
acceleration in a given external gravitational field. Therefore, for weak 
gravitational fields, a freely falling observer can be considered an inertial
frame. As anticipated, this is a consequence of starting
from an originally Lorentz-invariant theory and where all possible deviations 
represent just different aspects of the same physical phenomenon: gravitation. 
Freely falling in weak gravitational fields, 
is just a way to recover approximate Lorentz-covariance.

For this reason, the basic question about the validity of our picture 
reduces to the possibility
to distinguish between general relativity and {\it any} theory that 
incorporates the Equivalence Principle, 
by performing experiments 
to an accuracy ${\cal O}(G)$. More precisely, 
is it possible to explain the three classical
experimental tests (gravitational red-shift,
deflection of light and precession of perihelia)
without necessarily introducing the concept of a non-flat metric 
determined from the energy-momentum tensor by solving the
field equations with suitable boundary conditions ? If this is true,
classical general relativity cannot 
be considered the only possible description of gravity. 

Now, in the case of a centrally symmetric field 
both the gravitational red-shift 
 and the {\it correct} value for the
deflection of light were obtained by Schiff \cite{schiff}, long 
time ago by simple use of the Equivalence Principle and Lorentz transformations.
As a consequence, at the present, only
the precession of perihelia can be considered to depend on the
full details of Einstein theory, i.e. on the solution of the field equations 
represented by the Schwarzschild metric
\BE
  ds^2= 
c^2 dt^2 \left[1- {{2 GM}\over{c^2r}}\right] -
                 {{dr^2} \over{1 - {{2 GM}\over{c^2r }} } } - dl^2
\EE
where 
\BE
dl^2= r^2\left[d\theta^2+ \sin^2\theta d\varphi^2\right]
\EE
On the other hand, the
Equivalence Principle is a weak-field property and for the case of 
the perihelia the relevant gravitational effects are ${\cal O}(10^{-2})$ 
weaker than for
the deflection of light. This suggests that the same technique 
should also work in this case.
Due to the importance of the issue, we shall describe the proof in detail.

Let us start by considering the meaning of 
Eq.(6.1). In general relativity, this is a solution of the field equations
with flat-space boundary conditions at infinity. On the other hand, 
without specifying the units of length and time 
$dr$, $dt$ in (6.1) one cannot understand 
the physical interpretation of the reference frame 
where the precession is actually measured \cite{pauli}. 

In the case of the gravitational field of a large 
mass $M$ (e.g the sun) let us consider the set
of bound observers $O(i)$'s, freely falling along Keplerian
orbits $r\sim r(i)$,  and denote their 
space-time units ($d t(i), dr(i), dl$). 
To transfer the informations at spatial infinity, we can use
a freely falling observer $K_o$ with zero total
energy in the gravitational field. This can be considered as moving
with a radial `escape' velocity with respect to the $O(i)$'s
\BE
           v^2(i)= {{2 GM}\over{r(i)}}
\EE
and, up to a rotation, coincides asymptotically with the observers at rest at
spatial infinity in the condition of vanishing gravitational field.

We shall restrict to a weak-field condition so that the 
corrections to the classical theory
can be evaluated by considering circular orbits of radius
$r(i)$.  The order of the $O(i)$'s  
is such that $r(i)< r(i+1)$ and we assume 
$ {{2 GM}\over{c^2r(1)}} 
\ll 1$. For instance, for the 
sun, the value $r(1) \sim R_{\rm sun}$ gives
${{2 GM}\over{c^2r(1)}} \sim 4 \cdot 10^{-6}$.

To leading order, the relation between the $O(i)$'s and 
$K_o$ is a Lorentz-transformation with velocity (6.3) so that
the set of flat metrics 
\BE
      ds^2(i)= c^2dt^2(i) - dr^2(i) - dl^2
\EE
implies
\BE
  ds^2_o= 
c^2 dt^2(i) \left[1- {{2 GM}\over{c^2r(i)}}\right] -
                 {{dr^2(i)} \over{1 - {{2 GM}\over{c^2r(i) }} } } - dl^2
\EE
Namely, if $K_o$ wants to measure the space-time interval between two events with
radial components infinitesimally close to $r=r(i)$, by using
the space-time units of the corresponding $O(i)$, obtains Eq.(6.5). 
Notice, however, that the relation between the $O(i)$'s and $K_o$ is
a Lorentz transformation. Therefore, Eqs.(6.4) and (6.5) do {\it not} refer
to the same pair of events.

To address the problem of perihelia we observe that
the Keplerian orbits are computed
by assuming the validity of the Galilean tranformation $t'=t$, thus giving 
an absolute
meaning to angular velocities. This is no longer true since we know that
$ {{dl}\over{dt(i)}} \neq  {{dl}\over{dt_o}}$. 
The difference can be treated as a small perturbation to the Keplerian orbit
\BE
\delta U = - {{L^2}\over{2mr^2}} {{2GM}\over{c^2r}} \equiv {{\gamma}\over{r^3}}
\EE
where $L$ and $m$ 
denote the angular momentum and the mass of the 
body in the bound orbit. To first-order in $\delta U$, we can 
use the result of \cite{exercise}
\BE
        \Delta \varphi= -{{6\pi\gamma}\over{GMm p^2}}
\EE 
where $p^2\equiv a^2(1-e^2)^2$, $a$ and $e$ being
the parameters of the bound orbit. 
Therefore, replacing the value of the angular momentum
\BE
        L^2= {{4m^2 \pi^2}\over{T^2}} a^4(1-e^2)
\EE
we get the final expression
\BE
        \Delta \varphi= {{24 \pi^3 a^2} \over{T^2 c^2 (1-e^2)}}
\EE 
that, indeed, is the same expression as computed in general relativity.
For this reason, 
by following the original suggestion by Schiff \cite{schiff}, we 
conclude that,
to the present level of accuracy, all classical experimental tests of
general relativity would be fulfilled
in {\it any} theory that incorporates the Equivalence Principle. 

This result reflects the very general nature of 
the infinitesimal transformation to the rest frame of
a freely falling elevator. For instance, this can also
be implemented with a conformal (acceleration)
transformation of space and time  and a  tranformation of mass 
\cite{fulton} 
\BE
           m \to m [1 + \tilde{\phi}(x)]
\EE
that includes the gravitational energy. Eq.(6.10) would be
extremely natural in an approach where 
the gravitational potential is due to a
long-range fluctuation of the shifted Higgs field 
and suggests some considerations on the important
physical meaning of conformal transformations and the
basic assumption of a riemannian space-time in general relativity. 

Suppose we follow the basic idea underlying the Equivalence Principle, namely
a sequence of infinitesimal acceleration tranformations
to remove the effects of a weak gravitational field. 
Which is the final space-time structure ? This is an interesting question since
{\it we are} in a freely falling reference frame and, therefore, we want
to understand the properties of space-time also from this point of view.
Since we start from Minkowski space-time, an
acceleration transformation is naturally defined as an element
of the 15-parameter group $C_o$ in terms of a constant 4-vector
$\kappa_{\mu}$ 
\BE
 x'_{\mu}= {{x_{\mu} + \kappa_{\mu}x^2 }\over{ 1+ 2(\kappa \cdot x) + 
\kappa^2x^2 }}
\EE
This is singular at some values of $x_{\mu}$.
For instance, in the case $\kappa_{\mu}=(0;0,0,{{g}\over{2c^2}})$
the singularity occurs for $t={{2c}\over{g}}|1-{{g}\over{2c^2}}z|$. 
The point is
that only the product of the Poincare' group with scaling tranformations, the 
Weyl group, operate {\it globally} on ${\cal M}_o$ \cite{segal2}. Thus, if $X$ is the
generator of an infinitesimal acceleration
transformation (that is not in the Weyl group) and $p$ is a 
given point of 
${\cal M}_o$, $e^{(sX)}p$ is well defined for a sufficiently
small value of $s$ (whose range depends on $p$) but for no value of $s$ is
$e^{(sX)}$ defined throughout ${\cal M}_o$. 

Since the local gravitational field is not a constant, i.e.
$\kappa_{\mu}=\kappa_{\mu}(x)$, 
the singularity has not a real physical meaning but represents
a signal that the flat Minkowski space-time is carried 
out of itself into a larger covering space with typical 
local curvature 
$\sim {{c^2}\over{|g(x)|}}$. In general, 
conformal transformations lead to a covering space that is {\it not} a
riemannian space but a Weyl space \cite{fulton}.
The same problem,
in the framework of induced-gravity theories \cite{fuji,zee,adler}, means that
one should compute the effective lagrangian in 
a general background metric where
besides a symmetric tensor $g_{\mu \nu}$, there is a vector
$\kappa_{\mu}(x)$ at each point so that the Weyl connection is different
from the Christoffel symbol.

The crucial point, however, is that $\kappa_{\mu}$ is a gradient, 
since the gravitational acceleration field is
$\kappa_{\mu}=\partial_{\mu} \tilde{\phi}$, so that 
the Weyl space is {\it equivalent} to a riemannian
space \cite{fulton}. In this case, i.e. when
the origin of the gravitational acceleration is due to
a scalar potential, the 
basic assumption of a riemannian space becomes consistent with the intuitive
indications obtained from acceleration transformations. Therefore, 
the weak-field effective lagrangian in a background
space with arbitrary metric tensor $g_{\mu \nu}$ and Weyl connection 
$\kappa_{\mu}=\partial_{\mu} \tilde{\phi}$ 
will always reproduce Einstein field equations, for a suitable 
choice of the energy-momentum tensor. Here `suitable' means that the 
Minkowski-space $T_{\mu \nu}$ is, in general, 
modified for $\tilde{\phi}$-dependent terms and
that the definition of the energy-momentum tensor relevant for Einstein 
field equations does not contain any {\it large} cosmological term
from spontaneous symmetry breaking. The latter is an
obvious consequence of generating
the theory in curved space-time with a series of 
conformal transformations from flat space.
 
On the other hand, quite independently of conformal transformations, 
a very intuitive argument to understand why a large term as
$g_{\mu \nu} {\cal E}(n_v)$
cannot enter Einstein field equations
is the following. In Einstein's original
picture, all forms of energy and matter contribute to the 
space-time curvature. However, phions, although possessing 
an inertial mass, have no `gravitational mass' and cannot generate any 
curvature. This statement represents 
the geometrical counterpart of a non-Lorentz
covariant energy spectrum $\tilde{E}(k)$ 
for $k \to 0$, responsible for the 
instantaneous nature of the gravitational force. From this point of view, the
scalar condensate is, for gravitational phenomena, a real preferred frame and
can be considered the quantum realization of the 
old-fashioned {\it weightless} aether. Notice that
the idea of a {\it quantum} aether, of the type that can be generated from the
ground state of a quantum field theory, was considered by Dirac long time ago
\cite{dirac}. In this case, Dirac's 
aether velocity field $u_{\mu}$ coincides with $\kappa_{\mu}$ 
and represents the four-dimensional analogue 
of the superfluid velocity flow (5.14).

Finally, there seems to be good experimental evidence for
a {\it small} cosmological term \cite{roos}. Its
typical size is comparable with the ordinary-matter contribution and 
the combination of the two effects gives precisely a spatially-flat universe.
Outside of our framework, i.e. without giving a
special role to conformally-flat space, 
it would be very hard to understand this
result. On the other hand, by following 
the picture where the space-time curvature
is generated by a sequence of acceleration transformations on Minkowski space,
this small cosmological term is another way to introduce 
the cosmic acceleration field $g_o$ Eqs.(5.19)-(5.21) associated with the 
expansion of the universe.

If, in the end, it will turn out that 
the `preferred' \cite{wilc} metric structure $\bar{g}_{\mu \nu}$ 
of the universe requires the introduction 
of a cosmological term,  this will provide
an effective graviton mass term in Einstein weak-field 
equations for $g_{\mu \nu} \sim \bar{g}_{\mu \nu} + h_{\mu \nu}$.
  At the same time, if this has 
to reflect the dynamical origin of gravity, 
it may correspond to a superluminal 
propagation of gravitational `waves' \cite{binary}. 
As anticipated in Sect.2, our point of view
is closely related to the violations of causality in general relativity with a 
cosmological constant \cite{godel}.

\setcounter{equation}{0}
\section{Summary and concluding remarks}

In this paper we have presented a simple physical picture where
Newtonian gravity arises as a long-wavelength excitation of the scalar
condensate inducing spontaneous symmetry breaking. 
Our proposal represents the most natural interpretation of an important
phenomenon that has been missed so far:
the gap-less mode of the (singlet) Higgs field. 
Its existence is a direct consequence of Bose-Einstein 
condensation and has to be taken into account {\it anyway} (namely, 
how can we interpret it without gravity ?).

We emphasize that our
main result in Eq.(5.4) depends only on the Riemann-Lebesgue theorem on Fourier 
transforms \cite{goldberg} and 
two very general properties of the excitation spectrum. Namely, the
 `diluteness'  condition Eq.(4.16) (that leads to the `sound-wave' shape in
Eqs.(4.11) and (4.13) for $k \to 0$) and
the Lorentz-covariance for large $k$ (that leads to Eq.(4.10)). 
These features are expected to occur in
{\it any} description of spontaneous symmetry
breaking in terms of a weakly coupled Bose field and
depend on the weakly first-order nature of the
phase transition. This occurs for a very small but non-vanishing value of 
the phion mass $m$ so that
there is a non-relativistic regime $k \ll m$ where
the scalar condensate responds with phase-coherence. As discussed in detail in
Sects.3 and 4, this result reflects the existence 
of an ultimate ultraviolet cutoff responsible for the
deviations from an exactly Lorentz-covariant spectrum for $k \to 0$ in the 
broken phase.

A more complete
description of gravitational phenomena requires the detailed form of the energy
spectrum $\tilde{E}(k)$ and, in particular, the precise knowledge
of the phion mass $m$.  Deviations from the Newton potential are expected
at typical distances $r\sim 1/m$ and could, eventually, be
detected in the next generation of precise `fifth-force' experiments 
\cite{price}. We emphasize that these deviations from the $1/r$ law can 
change the description of the gravitational clustering of matter and are
essential to understand whether (or not) gravity remains in a weak-field
regime for a large gravitational mass. 
In our description, where gravity is the remnant of an almost `trivial' 
theory, this would be the most natural conclusion.

Finally, our description of 
the origin of gravity from the scalar condensate leads to the simple
hydrodynamical picture outlined in Eqs.(5.13)-(5.18). 
This provides a clue to the peculiar modification of Newtonian
gravity \cite{milgrom} that solves the experimental mass 
discrepancy in many galactic systems and
represents a completely new approach to the problem of dark matter. 

We emphasize that, our description of gravity, although predicting new 
phenomena, is not logically
in contradiction with general relativity, at least in an obvious way.
This can be understood by realizing
that two main outcomes of our picture, the Equivalence Principle and the 
`Mach's Principle' view of inertia were two basic ingredients at the origin 
of Einstein theory \cite{einstein}. For this reason, the classical
tests of general relativity in weak gravitational field
are fulfilled as in {\it any} theory incorporating the Equivalence
Principle. Further, the
special role of infinitesimal conformal 
transformations to implement the transition to the rest-frame of a freely falling
elevator suggests 
 that Einstein equations may represent the effective
weak-field approximation
of a theory generated from flat space with a sequence of
conformal transformations. This can easily explain the absence of a {\it large}
cosmological term in Einstein field equations.

We stress that
the apparently `trivial' nature of $\lambda\Phi^4$ theories in four 
space-time dimensions should not induce to overlook the possibility
that gravity can arise as a gap-less mode of the Higgs field.
Indeed, our description is only possible if one assumes the existence of 
an ultimate ultraviolet cutoff so that the natural formulation of the
theory is on the lattice and `triviality' is never complete.
In this case, however, the existence of a non-trivial infrared behaviour 
in the broken phase is not surprising due to
the equivalence of low-temperature Ising models with highly
non-local membrane models on the dual lattice \cite{gliozzi}
 whose continuous
limit is some version of the Kalb-Ramond \cite{kalb}
model. Thus, in the end, a Higgs-like description of gravity
 may turn out to be equivalent, at some scale, to a 
Feynman-Wheeler theory of {\it strings}, 
as electromagnetism for point particles \cite{hoyle}.
At the same time, the basic idea that
one deals with the {\it same} theory should allow to replace
a description with its `dual' picture when better suited to provide
an intuitive physical insight. 

To conclude, we want to mention another implication of the 
scalar condensate at the astronomical level. As anticipated, 
our picture provides a natural solution of the
so-called `hierarchy-problem'. This depends on 
the {\it infinitesimally weak} first-order nature of the phase
transition: in units of the Fermi scale, 
 $M_h\sim G_F^{-1/2}$, the Planck 
scale $G^{-1/2}$ would diverge for a vanishing phion mass $m$. 
These scales are hierarchically 
related through the large number $\sqrt{\eta} \sim 10^{17}$ which is the only
manifestation of an ultimate ultraviolet cutoff. In this sense, spontaneous
symmetry breaking in $(\lambda\Phi^4)_4$ theory represents an
approximately scale-invariant 
phenomenon and it is conceivable that powers of
the `replica-factor' $10^{17}$ will further show up in a natural way
(notice that in our picture
the factor $\eta^{1/6}={\cal O}(10^{5})$ also appears, 
see Sect.4 and fig.4 of ref.\cite{mech}).  For this reason, after the 
scales $m\sim 10^{-5}$ eV and $\sqrt{\eta} m \sim 10^{2}$ GeV, 
we would expect the scale 
${{m}\over{\sqrt{\eta} }} \sim 10^{-22}~ {\rm eV}$ 
to play a role \cite{qed} 
at the astronomical level in connection with a length
\BE
l \sim {{\sqrt{\eta} }\over{m}} \sim 10^{17}~{\rm cm}.
\EE
To this end, we observe that, in the field of a large mass as the
sun, the `MOND' regime \cite{milgrom} mentioned in Sect.5
corresponds precisely to the length scale $l$ , namely 
\BE
  g_o \sim {{GM_{\rm sun} }\over{ l^2}} \sim 
10^{-8}~{\rm cm}~{\rm sec}^{-2}
\EE
As anticipated, 
in our picture this also corresponds to a situation where the phion density
$n$ differs substantially 
from its average value (5.16) related to the Newton constant and the
Poisson equation (5.12) and the equivalent condition of a constant-density
potential flow Eqs.(5.14)-(5.15) cease to be valid. This means 
that beyond $l$ we enter in a `$G$-variable' theory \cite{path}.

Even at distances $r \ll l$ the effects of $g_o$ may, however, be observable
if there is a Hubble flow over solar-system 
scales. As discussed in Sect.5, in fact, there would be a
 cosmological velocity field $\langle {\bf{u}}_s \rangle$ associated with the
continuous creation of phions in an expanding universe. The basic 
idea of a Hubble
flow over small scales has been recently re-proposed \cite{rosales,guru} in connection
with the observation by Anderson {\it et al} \cite{pioneer} of an
anomalous acceleration $g_{\rm anom} \sim g_o$ 
from the Pioneer data 
at the border of the solar system ($r \sim 10^{15}$ cm). 
Since this interpretation has been shown \cite{guru} to be consistent
with planetary and even geological data, 
(the unexplained part of) the
Pioneer anomaly may represent the first evidence of a fundamental phenomenon
related to the true dynamical origin of inertia and gravity.

Finally, at very large distances $r >> l$ 
one has to use the more general Eqs.(5.17)-(5.18) 
to determine the gravitational field
for a given distribution of matter $\sigma(x)$. 
In this case, the properties of
galactic systems require strong departures from Newtonian dynamics and
the gravitational acceleration is found
$g \sim 1/r$ rather than $1/r^2$ \cite{milgrom}. A full understanding of
this result
requires to {\it predict} the form of the function $f$ in (5.18) by evaluating
those residual self-interaction effects in the 
scalar condensate responsible for the transition from the constant-density
regime described by
Eqs.(5.12), (5.15) to the more general Eqs.(5.17), (5.18). The idea of
explaining the mass discrepancy without any form of `dark matter' is not
unconceivable. Indeed, the typical values for the visible masses of 
gravitating matter
\BE
M_{\rm galaxy} \sim 10^{11} M_{\rm sun} \sim \eta^{1/3} M_{\rm sun}
\EE
and 
\BE
\rho_{\rm galaxy} \sim 10^4 {\rm parsec} \sim \eta^{1/6} l 
\EE
lead to the same value as in Eq.(7.2)
\BE
  g_o \sim {{GM_{\rm galaxy} }\over{ \rho^2_{\rm galaxy} }} \sim 
10^{-8}~{\rm cm}~{\rm sec}^{-2}
\EE
Similarly, the values ( U= visible universe)
\BE
M_{\rm U } \sim  \eta^{1/3} M_{\rm galaxy} \sim \eta^{2/3} M_{\rm sun}
\EE
and
\BE
\rho_{\rm U} \sim 10^{10}~ {\rm light~years} \sim \eta^{1/3} l 
\EE
give again
\BE
  g_o \sim {{GM_{\rm U } }\over{ \rho^2_{\rm U} }} \sim 
10^{-8}~{\rm cm}~{\rm sec}^{-2}
\EE
This suggests that the properties of the
scalar condensate can indeed play an essential role 
to understand the steps in the cosmological 
hierarchy \cite{devau,baryshev2}.
\vskip 20 pt

{\bf Note added in proof}

After finishing this paper I have become aware of 
a recent preprint by F. Ferrer and J. A. Grifols, {\it Effects of 
Bose-Einstein condensation on forces among bodies sitting in a boson heat bath},
 hep-ph/0001185.  In this paper, the presence of
a long range $1/r$ potential in connection with the Bose Einstein condensation
of a massive
scalar field is also pointed out. I thank P. M. Stevenson for this information.

\vfill
\eject
\centerline{\large{ \bf {Figure Caption} } }
\vskip 20 pt
\par {\bf Fig.1} ~ A qualitative pictorial representation of the energy spectrum 
$\tilde{E}(k)$. The actual relative sizes are such that
the continuous matching inside the shaded blob is
at a value $k \sim m ={\cal O}(10^{-17})$ in units of $M_h$.


\begin{thebibliography} {99}
\bibitem{universal}
By `universal' we mean that in dilute Bose systems the excitation spectrum 
for $ k \to 0$ can be described in terms of two parameters: the particle density
$n$ and the physical scattering length $a$. This latter parameter summarizes
in its physical value all relevant features 
of the short-range 2-body interaction, see ref.\cite{huang}.
\bibitem{bondi}
This picture, closely 
rensembles some features of Steady-State cosmology \cite{steady}. In our case,
it is a consequence of spontaneous symmetry breaking
in an expanding universe.
\bibitem{remark}
Strictly speaking, 
this intuitive mechanism to understand the Mach Principle neglects the 
continuous creation of phions mentioned in sect.5 in connection with the 
expansion of the universe. In this sense, inertia would vanish only in the absence
of the compensating effect that tends to establish the value $n=n_v$ everywhere.
This reduces the dependence
of inertia on the distribution of gravitational sources.
\bibitem{binary}
Just for this reason, the observed slowing down of binary systems
cannot be considered at the same level as the
direct detection of gravitational waves propagating at the speed of light. 
We stress that an energy transfer from fastly
rotating objects to the scalar condensate 
is not in contradiction with the mentioned properties of the vacuum. Indeed,
as anticipated in Sect.4, just
in the case of rotations, superfluid systems can be excited for values of the
angular velocity that are several orders of magnitude
 smaller than the sound velocity \cite{feynman}. Admittedly,
we have not worked out any quantitative estimate of this possible reduction
effect.
\bibitem{qed}
 This is in analogy with QED where, besides
the electron Compton wavelength $1/m_e$, the classical electron radius
 ${{\alpha}\over{m_e}}$
and the Bohr radius ${{1}\over{\alpha m_e}}$ also appear.
\bibitem{path}
A change in the phion density at
$r\sim l$ is also suggested by the interpretation of $l$ as the
mean free path for condensed phions in the typical regime of density 
associated with Eq.(5.16). Indeed, as anticipated at the end of Sect.4,
the fraction of
phions that, in the absence of an external field, populate the states with 
$k \neq 0$, give rise to scattering processes and a small friction
in a superfluid flow. 
The number of collisions in a generic path of length $L$
due to the presence of a non-zero depletion (see Eqs.(4.16)-(4.18) ) is
${{N_{\rm coll} } \over{ L}} \sim 4 \pi a^2 n_D$ and
can be estimated
by using the values $am= {\cal O} (1/\eta)$, 
$n_D= {\cal O}(1/{\sqrt \eta}) \langle n \rangle$ and Eq.(5.16). In this way, 
we obtain a large, but finite, mean free path 
 $l \sim {{ \sqrt{\eta} }\over{ m }} 
={\cal O}(10^{17})~{\rm cm}$
\end{thebibliography}
\end{document}